\documentclass[showpacs,amssymb,preprint,preprintnumbers,nofootinbib,superscriptaddress]{revtex4}

\usepackage{amsmath}
\usepackage{graphicx}
\graphicspath{{Figures/}}
\usepackage{latexsym}
\usepackage{amsfonts}
\usepackage{url,hyperref}
\usepackage{bm}
\usepackage{textcomp}
\usepackage{color}
\usepackage{bbm}
\usepackage{slashed}
\usepackage{caption}
\usepackage{epstopdf}
\usepackage{subcaption}
\usepackage{placeins}
\usepackage{color}

\newcommand{\beq}{\begin{equation}}
\newcommand{\eeq}{\end{equation}}
\def\bea{\begin{eqnarray}}
\def\eea{\end{eqnarray}}

\begin{document}

\title{Master equations and Quasi-normal modes of spin-3/2 fields in Schwarzschild (A)dS black hole spacetimes}

\author{Chun-Hung~Chen}
\email[Email: ]{chun-hungc@nu.ac.th}
\affiliation{The Institute for Fundamental Study, Naresuan University, Phitsanulok 65000, Thailand.}

\author{Hing-Tong~Cho}
\email[Email: ]{htcho@mail.tku.edu.tw}
\affiliation{Department of Physics, Tamkang University, Tamsui District, New Taipei City, Taiwan 25137.}

\author{Alan~S.~Cornell}
\email[Email: ]{acornell@uj.ac.za}
\affiliation{Department of Physics, University of Johannesburg, PO Box 524, Auckland Park 2006, South Africa.}

\author{Gerhard~E.~Harmsen}
\email[Email: ]{gerhard.harmsen5@gmail.com}
\affiliation{Van Swinderen Institute, University of Groningen, 9747 AG, Groningen, The Netherlands.}

\begin{abstract}
In this work we consider spin-3/2 fields in Schwarzschild (A)dS black hole spacetimes. As this spacetime is different from the Ricci-flat cases, it is necessary to modify the covariant derivative to the supercovariant derivative in order to maintain the gauge symmetry, as noted in our earlier works, where this is done here by including terms related to the cosmological constant. Together with the eigenmodes for spin-3/2 fields on an $n$-sphere, we derive the master radial equations, which have effective potentials that in general include an explicitly imaginary part and energy dependence. We found that for the asymptotically AdS cases, the explicit imaginary dependence automatically disappears, because of the negative cosmological constant. We take this case as an example and obtain the quasi-normal modes by using the Horowitz-Hubeny method \cite{hh2000}.
\end{abstract}


\date{\today}

\maketitle

%
%

\section{Introduction}

\par Obtaining the master equations for black hole perturbation theory by considering different fields in general dimensional spherically symmetric black hole backgrounds is a well studied research topic, except for the spin-3/2 fields, where only a limited literature exists. This paper is the third part of a systematic study of spin-3/2 fields in higher dimensional Schwarzschild \cite{ccch2016} and Reissner-Nordstr\"{o}m black hole spacetimes \cite{ccch2018}, where we now seek to confirm the methods used for obtaining the radial equation for cases involving a cosmological constant, and to test how the behavior of the spin-3/2 field shall affect the quasi-normal modes (QNMs).

\par As in our previous works, we consider the Rarita-Schwinger equation \cite{rarsch} in the form which describes the behavior of a gravitino in supergravity theories \cite{dasfre,gripen}:
\begin{equation}\label{RSE}
\gamma^{\alpha\mu\nu}\nabla_{\mu}\psi_{\nu}=0\; ,
\end{equation}
where $\gamma^{\alpha\mu\nu}=\gamma^{[\alpha}\gamma^{\mu}\gamma^{\nu]}$ is the totally antisymmetric product of Dirac gamma matrices, $\nabla_{\mu}$ is the covariant derivative, and $\psi_{\nu}$ is a spinor-vector field. The Rarita-Schwinger equation in this form is convenient for generalizing into higher dimensional cases, however, the spinor-vector field $\psi_{\nu}$ will in general not be gauge invariant under the transformation
\begin{equation}\label{GT}
\psi_{\nu}'=\psi_{\nu}+\nabla_{\nu}\varphi \; ,
\end{equation}
where $\varphi$ is a gauge spinor. To maintain the gauge invariance of Eq. (\ref{RSE}) we have to replace the covariant derivative with a ``supercovariant derivative" which we derived previously in the Reissner-Nordstr\"{o}m case \cite{ccch2018}, and which we shall derive for the (A)dS spacetimes here. This becomes the essential difference for the study of spin-3/2 fields when compared with the study of the other spin fields in higher dimensional spherically symmetric spacetimes.

\par In this paper, we shall therefore generalize our previous works by including the cosmological constant in the Einstein field equations, which indicates that the black hole will not be located in an asymptotically flat spacetime but an (A)dS one. The master radial equations, and their effective potentials, shall be derived, where it is found that they shall have an explicit imaginary part, as well an energy dependence. As such, it shall be more convenient to choose the higher dimensional asymptotically AdS cases in pursuing our QNM analysis. Note that in this case the effective potentials behave like a confining box, which is a general expectation for perturbing fields in an asymptotically AdS background. This allows us to use the Horowitz-Hubeny method \cite{hh2000} with Dirichlet boundary conditions to study the QNMs. 

\par Recall that the black hole QNMs originally represented the cosmological black hole's ringing, but in the asymptotically AdS cases this may be slightly different, as is suggested by the AdS/CFT correspondence \cite{hh2000,bss2002,mn2002}. Note that some works have given a more precise comment on the QNMs under such an equivalence, by relating them to the poles of retarded Green's functions of gauge invariant operators in supersymmetric Yang-Mills theory \cite{ns2003}. Furthermore, many works on QNMs in asymptotically AdS cases have been done with bosonic perturbations \cite{kono2002,cl2001,ckl2003,wlm2004,whs2015}, Dirac perturbations \cite{gj2005,whj2017}, and large overtone spin-3/2 perturbations \cite{asv2014}. Our works have attempted to fill the gap in this literature, by developing formulae for spin-3/2 perturbations.

\par As such, we have structured this paper as follows: In the next section we define the ``supercovariant derivative" for the Rarita-Schwinger equation, Eq.~(\ref{RSE}), corresponding to the Schwarzschild-(A)dS background. In sec. \ref{sec:3} we obtain the master radial equations both for the ``non-Transverse and Traceless" (non-TT), and ``Transverse and Traceless" (TT) modes \cite{ccch2016, ccch2018}. We will then present some discussion about the properties of the effective potentials in sec. \ref{sec:4}, and in sec. \ref{sec:5} present the modified Horowitz-Hubeny method to fit the equations under investigation here, obtaining the spin-3/2 QNMs for the asymptotically AdS black hole background. Finally in sec. \ref{sec:6}, we present our conclusions and some related future directions for this work.

%
%

\section{Super covariant derivative}

\par In order to maintain gauge invariance of the spinor-vector field $\Psi_{\nu}$ in the Rarita-Schwinger equation, we have to rewrite the derivative in terms of a supercovariant derivative, which incorporates the gauge transformation, Eq.~(\ref{GT}), into Eq.~(\ref{RSE}). The condition shall therefore be
\begin{equation}\label{GIC}
\gamma^{\alpha\mu\nu}\left[\mathcal{D}_{\mu},\mathcal{D}_{\nu}\right]\varphi=0 \; ,
\end{equation}
where $\mathcal{D}_{\mu}=\nabla_{\mu}+a\sqrt{\Lambda}\gamma_{\mu}$, and $a$ is a yet to be determined factor, which we are required to solve to satisfy Eq.~(\ref{GIC}). Together with the relation $\nabla_{\mu}\gamma_{\nu}=\nabla_{\mu}e^{a}_{\nu}\gamma_{a}=0$, Eq.~(\ref{GIC}) becomes
\begin{eqnarray}
0&=&\gamma^{\alpha\mu\nu}\left[\mathcal{D}_{\mu},\mathcal{D}_{\nu}\right]\psi=\gamma^{\alpha\mu\nu}\left[\nabla_{\mu},\nabla_{\nu}\right]\psi +a^{2}\Lambda\gamma^{\alpha\mu\nu}\left[\gamma_{\mu},\gamma_{\nu}\right]\psi \nonumber\\
&=&\gamma^{\mu}G_{\mu}^{\ \alpha}\psi+a^{2}\Lambda\left(-2\left(D-2\right)\left(D-1\right)\gamma^{\alpha}\right)\psi \; ,
\end{eqnarray}
where $G_{\mu}^{\ \alpha}$ is the Einstein tensor. Together with the source free Einstein field equations, we found that if
\begin{equation}
a=\frac{i}{\sqrt{2\left(D-2\right)\left(D-1\right)}}\; ,
\end{equation}
then Eq.~(\ref{GIC}) will always be true. That is, if we rewrite the Rarita-Schwinger equation, Eq.~(\ref{RSE}), with the supercovariant derivative in Schwarzschild (A)dS spacetimes, then
\begin{eqnarray}\label{SCD}
0&=&\gamma^{\alpha\mu\nu}\mathcal{D}_{\mu}\psi_{\nu},\nonumber\\
&=&\gamma^{\alpha\mu\nu}\left(\nabla_{\mu}+i \sqrt{\frac{\Lambda}{2\left(D-2\right)\left(D-1\right)}}\,\gamma_{\mu}\right)\psi_{\nu} \; .
\end{eqnarray}

\par It is worth recalling that the supercovariant derivative from our previous work, Ref.~\cite{ccch2018}, arose for spin-3/2 fields in general dimensional Reissner-Nordstr\"{o}m cases because the spacetimes was no longer Ricci flat. Note that we do not need to have the supercovariant derivative in Ricci flat cases as Eq.~(\ref{GIC}) will be automatically satisfied \cite{ccch2016}. For the higher dimensional Reissner-Nordstr\"{o}m cases we had to introduce the supercovariant derivative, as in Eq.~(2.9) in Ref.~\cite{ccch2018}, to ensure the gauge invariance of the Rarita-Schwinger equation. This is modified for our current spacetime in Eq.~(\ref{SCD}), to reflect the presence of the cosmological constant. This is the essential difference when studying spin-3/2 perturbations in comparison with other field perturbations in spherically symmetric spacetimes. Note that the supercovariant derivative we have presented in Eq.~(\ref{SCD}), and our previous works, are consistent with the results of Ref.~\cite{lzy2014}.

%
%

\section{Equations of motion and master radial equations}\label{sec:3}

\par To obtain the master radial equations we begin by defining the line element that we will use. In this case it is given as
\begin{equation}
ds^{2}=-f(r)dt^{2}+\frac{1}{f(r)}dr^{2} + r^{2}d\Omega_{N}^{2} \; ,
\end{equation}
where
\begin{equation}
f(r)=1-\frac{2M}{r^{D-3}}-\frac{2\Lambda r^{2}}{(D-2)(D-1)}\; ,
\end{equation}
and $d\Omega_{N}^{2}$ is the metric of an $N$-sphere with $D=N+2$. Note that we will use the over bars to denote terms from the $d\Omega_{N}^{2}$ metric. For the Rarita-Schwinger equation, Eq.~(\ref{SCD}), in this spacetime, we may obtain two sets of master radial equations, the non-TT and TT modes. This comes from the separation of the angular part of the equations of motion by using the ``non-TT eigenmodes'' and ``TT eigenmodes'' on $S^{N}$. For details on the spinor-vector eigenmodes, as well as the choice of Dirac gamma matrices, and the calculation of the spin connections, we refer the reader to Refs.~\cite{ccch2016,ccch2018}.

\subsection{Radial equations with non-TT eigenmodes}

\par We proceed by taking the spin-3/2 fields in the form
\begin{eqnarray}\label{nttSVF}
\psi_{t}=\phi_{t}\otimes\bar{\psi}_{(\lambda)} \,\,\, \text{and} \,\,\, \psi_{r}=\phi_{r}\otimes\bar{\psi}_{(\lambda)}\; . \nonumber\\
\psi_{\theta_{i}}=\phi_{\theta}^{(1)}\otimes\bar{\nabla}_{\theta_{i}}\bar{\psi}_{(\lambda)} + \phi_{\theta}^{(2)}\otimes\bar{\gamma}_{\theta_{i}}\bar{\psi}_{(\lambda)}\; ,
\end{eqnarray}
where $\bar{\psi}_{(\lambda)}$ is an eigenspinor on $S^{N}$, with eigenvalues $i\bar{\lambda}$, and $\phi_{t}$, $\phi_{r}$, $\phi_{\theta}^{(1)}$ and $\phi_{\theta}^{(2)}$ are functions of $r$ and $t$ only, and  behave as 2-spinors. The eigenvalues $\bar{\lambda}$ are given by $\bar{\lambda}=j+(D-3)/2$, where $j=3/2,\,5/2,\,7/2,\,...$ \cite{ccch2016}. We use the Weyl gauge, that is $\phi_{t}=0$, in which case out of the four equations of motion, only three are independent. After simplifying, they can be written as:
\begin{eqnarray}
0&=&\left(i\bar{\lambda} \partial_{r} + (D-3)\frac{i\bar{\lambda}}{2r}-\frac{(D-2)(D-3)}{4r\sqrt{f}}i\sigma^{3}-\frac{\bar{\lambda}}{\sqrt{f}}\sqrt{\frac{\Lambda(D-2)}{2(D-1)}}\sigma^{2}\right)\phi^{(1)}_{\theta} \nonumber\\
&&+\left((D-2)\partial_{r} + (D-3)\frac{i\bar{\lambda}}{r\sqrt{f}}i\sigma^{3}+\frac{(D-2)(D-3)}{2r}+(D-2)\frac{i}{\sqrt{f}}\sqrt{\frac{\Lambda(D-2)}{2(D-1)}}\sigma^{2} \right)\phi^{(2)}_{\theta} \nonumber\\
&&-\left(i\bar{\lambda}+\frac{D-2}{2}\sqrt{f}\,i\sigma^{3}+ir\sqrt{\frac{\Lambda(D-2)}{2(D-1)}}\sigma^{1} \right)\phi_{r}\; ,\nonumber\\
0&=&\left(-\frac{i\bar{\lambda}}{\sqrt{f}}\partial_{t} + (D-3)\frac{i\bar{\lambda}\sqrt{f}}{2r}\sigma^{1} + \frac{i\bar{\lambda}f'}{4\sqrt{f}}\sigma^{1} -\frac{(D-2)(D-3)}{4r}\sigma^{2}-\bar{\lambda}\sqrt{\frac{\Lambda(D-2)}{2(D-1)}}\, i\sigma^{3}\right)\phi^{(1)}\nonumber\\
&&+\Big(-\frac{D-2}{\sqrt{f}}\,\partial_{t}+\frac{(D-3)(D-2)}{2r}\sqrt{f}\,\sigma^{1} + \frac{(D-2)f'}{4\sqrt{f}}\sigma^{1}+(D-3)\frac{i\bar{\lambda}}{r}\sigma^{2}\nonumber\\
&&-(D-2)\sqrt{\frac{\Lambda(D-2)}{2(D-1)}}\,\sigma^{3} \Big)\phi^{(2)}_{\theta}\; ,\nonumber 
\end{eqnarray}
\begin{eqnarray}
0&=&\left(\frac{i}{r\sqrt{f}}\sigma^{3}\partial_{t} + \frac{\sqrt{f}}{r}\sigma^{2}\partial_{r} + \frac{f'}{4r\sqrt{f}}\sigma^{2} + (D-4)\frac{\sqrt{f}}{2r^{2}}\sigma^{2} + \frac{i}{r}\sqrt{\frac{\Lambda(D-2)}{2(D-1)}} \right)\phi^{(1)}_{\theta}\nonumber\\
&&-\frac{D-4}{r^{2}}\sigma^{1}\phi^{(2)}_{\theta} - \frac{\sqrt{f}}{r}\sigma^{2}\phi_{r}\; .\label{EQ:nnTTEoM}
\end{eqnarray}
However, the variables $\phi_{r}$, $\phi_{\theta}^{(1)}$ and $\phi_{\theta}^{(2)}$ are not gauge invariant functions. As such, and as we have done previously, we will need to construct gauge invariant variables. From the gauge transformation in Eq.~(\ref{GT}), 
$$\psi_{\theta_{i}}^{'}=\psi_{\theta_{i}}+\nabla_{\theta_{i}}\varphi$$
\begin{eqnarray}
&\Rightarrow&\phi_{\theta}^{'(1)}\otimes\bar{\nabla}_{\theta_{i}}\bar{\psi}_{(\lambda)}+\phi_{\theta}^{'(2)}\otimes\bar{\gamma}_{\theta_{i}}\bar{\psi}_{(\lambda)}\nonumber\\
&&\hspace{0.5cm}=\left(\phi^{(1)}_{\theta}+\phi \right)\otimes\bar{\nabla}_{\theta_{i}}\bar{\psi}_{(\lambda)} + \left[\phi^{(2)}_{\theta} + \left(\frac{\sqrt{f}}{2}i\sigma^{3} + \frac{ir}{D-2}\sqrt{\frac{\Lambda(D-2)}{2(D-1)}}\,\sigma^{1}  \right)\phi\right]\otimes \bar{\gamma}_{\theta_{i}}\bar{\psi}_{(\lambda)}\; . \nonumber \\
&& \hspace{0.5cm}
\end{eqnarray}
Hence, a gauge invariant variable can be defined as
\begin{equation}
\Phi = - \left(\frac{\sqrt{f}}{2}i\sigma^{3}+\frac{ir}{D-2}\sqrt{\frac{\Lambda(D-2)}{2(D-1)}}\,\sigma^{1} \right)\phi^{(1)}_{\theta}+\phi^{(2)}_{\theta}\; .
\end{equation}
We now use this gauge invariant variable in Eq.~(\ref{EQ:nnTTEoM}), and simplify to obtain the following equation of motion for $\Phi$,
\begin{eqnarray}\label{EQ:CoupledEq}
\left(i\bar{\lambda} + \frac{D-2}{2}\sqrt{f}\,i\sigma^{3} \right. &-& \left. ir\sqrt{\frac{\Lambda(D-2)}{2(D-1)}}\,\sigma^{1}\right) \times \nonumber \\
\left[\sigma^{1}\partial_{t} -\frac{(D-3)f}{2r}- \frac{f'}{4}\right. &-& \left. i\bar{\lambda}\,\frac{D-3}{D-2}\frac{\sqrt{f}}{r}i\sigma^{3}
 - i\sqrt{\frac{\Lambda f(D-2)}{2(D-1)}}\,\sigma^{2}\right]\Phi\nonumber\\
&=&\left(i\bar{\lambda}-\frac{D-2}{2}\sqrt{f}\,i\sigma^{3}-ir\sqrt{\frac{\Lambda(D-2)}{2(D-1)}}\,\sigma^{1}\right) \times \\
&&\left[f\partial_{r}+i\bar{\lambda}\,\frac{\sqrt{f}}{(D-2)r}\,i\sigma^{3}+\frac{(2D-7)f}{2r}
+\frac{2i}{D-2}\sqrt{\frac{\Lambda f (D-2)}{2(D-1)}}\,\sigma^{2}\right]\Phi\; . \nonumber
\end{eqnarray}
If we take 
$$\Psi = \left(i\bar{\lambda} - \frac{D-2}{2}\sqrt{f}\,i\sigma^{3}-ir\sqrt{\frac{\Lambda(D-2)}{2(D-1)}}\,\sigma^{1}\right)\Phi \; ,$$ 
then we can rewrite the above equations as
\begin{equation}
f\partial_{r}\Psi + \left(\mathcal{A}+\mathcal{B}i\sigma^{3}+\mathcal{D}\sigma^{2}\right)\Psi = \sigma^{1}\partial_{t}\Psi \; ,
\end{equation}
where
\begin{eqnarray}\label{NTTCOE}
\mathcal{A} &=& \frac{1}{-\bar{\lambda}^{2}+ \frac{(D-2)^{2}}{4}f + r^{2}\frac{\Lambda(D-2)}{2(D-1)}}\Bigg[-\bar{\lambda}^{2}\left(\frac{f'}{4}+\frac{D-4}{2r}f \right) + \frac{(D-2)^{2}}{4}f\left(-\frac{3}{4}f'+\frac{D-4}{2r}f \right)\; ,\nonumber\\
&&\hspace{6cm} - r^{2}\,\frac{\Lambda(D-2)}{2(D-1)}\left( -\frac{f'}{4}-\frac{(D-8)f}{2r}\right)\Bigg]\; ,\nonumber\\
\mathcal{B} &=& \frac{i\bar{\lambda}\sqrt{f}}{r}\left[1 + \frac{1}{-\bar{\lambda}^{2}+ \frac{(D-2)^{2}}{4}f + r^{2}\frac{\Lambda(D-2)}{2(D-1)}}\left( \frac{(D-2)(D-3)M}{r^{D-3}}\right) \right]\; ,\nonumber\\
\mathcal{D}&=&-i\sqrt{\frac{\Lambda f (D-2)}{2(D-1)}}\left[\frac{D-4}{D-2} + \frac{1}{-\bar{\lambda}^{2}+ \frac{(D-2)^{2}}{4}f 
+ r^{2}\frac{\Lambda(D-2)}{2(D-1)}}\left( \frac{(D-2)(D-3)M}{r^{D-3}}\right) \right]\; .
\end{eqnarray}
By using a further transformation $\Psi=\mathcal{K}(r)\widetilde{\Psi}$, we can remove the $\mathcal{A}$ term provided $\mathcal{K}(r)$ satisfies the differential equation
\begin{equation}
\frac{f}{\mathcal{K}}\frac{d\mathcal{K}}{dr}+\mathcal{A}=0\; .
\end{equation}
We are then left with an equation of the form
\begin{equation}\label{EQ:EoM1}
f\partial_{r}\widetilde{\Psi}+\left(\mathcal{B}i\sigma^{3}+\mathcal{D}\sigma^{2} \right)\widetilde{\Psi} = \sigma^{1}\partial_{t}\widetilde{\Psi}\; .
\end{equation}
Finally, we can separate the spinor $\widetilde{\Psi}$ into its components by making the choice
\begin{equation}
    \widetilde{\Psi} = \left[\sin\left(\frac{\theta}{2} \right)\sigma^{3}+\cos\left(\frac{\theta}{2} \right)\sigma^{2} \right]e^{-i\omega t}
    \begin{pmatrix}
    \phi_{1}\\
    \phi_{2}\\
    \end{pmatrix} \; ,\label{Psitilde}
\end{equation}
where the $\phi_{1,2}$ are functions of the radial coordinate only, and
\begin{equation}
    \theta = \tan^{-1}\left(\frac{-\mathcal{D}}{i\mathcal{B}}\right)\; .
\end{equation}
Note that a similar transformation was introduced for obtaining the radial equation of a massive Dirac particle in a 4-dimensional Schwarzschild black hole spacetime \cite{cho2003}.
Substituting the expression in Eq.~(\ref{Psitilde}) into Eq.~(\ref{EQ:EoM1}) and simplifying, we get
\begin{equation}
    \left(f\partial_{r} + \frac{f}{2}\left[\frac{\partial}{\partial r}\left(\frac{-\mathcal{D}}{i\mathcal{B}}\right)\right]\left(\frac{\mathcal{B}^{2}}{\mathcal{B}^{2}-\mathcal{D}^{2}} \right)i\sigma^{1} + \sqrt{\mathcal{D}^{2}-\mathcal{B}^{2}}\,\sigma^{3} \right)
    \begin{pmatrix}
    \phi_{1}\\
    \phi_{2}\\
    \end{pmatrix}=
    i\omega\sigma^{1}
    \begin{pmatrix}
    \phi_{1}\\
    \phi_{2}\\
    \end{pmatrix} \; ,
\end{equation}
which can be expanded to give
\begin{equation}
    \begin{aligned}\label{EQ:AdSEoM2}
    \left(\partial_{r_{*}} + W \right)\phi_{1}=i\omega\phi_{2}\; , \\
    \left(\partial_{r_{*}} - W \right)\phi_{2}=i\omega\phi_{1}\; . \\
    \end{aligned}
\end{equation}
Note that $\partial_{r_{*}}=\mathcal{F}\partial_{r}$ is the tortoise coordinate and
\begin{equation}\label{NTTSP}
\begin{aligned}
   \mathcal{F} = f\left[1 + \frac{f}{2\omega}\left(\frac{\partial}{\partial r}\frac{\mathcal{D}}{i\mathcal{B}} \right)\left(\frac{\mathcal{B}^{2}}{\mathcal{B}^{2}-\mathcal{D}^{2}} \right) \right]^{-1}\; ,\\
    W = \sqrt{\mathcal{D}^{2}-\mathcal{B}^{2}}\left[1 + \frac{f}{2\omega}\left(\frac{\partial}{\partial r}\frac{\mathcal{D}}{i\mathcal{B}} \right)\left(\frac{\mathcal{B}^{2}}{\mathcal{B}^{2}-\mathcal{D}^{2}} \right) \right]^{-1}\; .
\end{aligned}
\end{equation}
After decoupling Eqs.~(\ref{EQ:AdSEoM2}), we obtain
\begin{equation}\label{NTTRD}
    \left(\partial_{r_{*}}^{2} +\omega^{2}- V_{eff}\right)\phi_{1,2}=0 \; ,
\end{equation}
where the effective potential is
\begin{equation}\label{NTTEFV}
V_{eff}=\mp \partial_{r_{*}}W+W^{2} \; .
\end{equation}

\par There is a notable difference here for the effective potential from previous cases studied in Refs.~\cite{ccch2016, ccch2018}, in that the potential has an imaginary component, and a dependence on the QNM $\omega$. Note that in Refs.~\cite{ccch2016, ccch2018} the potential relied only on the radial coordinate. The presence of the $\omega$ does mean that the potential function, in its current form, is not guaranteed to be a real valued function. The function becomes real by taking the limit $\Lambda\rightarrow 0$ (removing the effect of the cosmological constant), in which the potential returns to the Schwarzschild effective potential case. To have a more complete understanding of the effective potential, including how it is affected by the cosmological constant, we will present some of its characteristics in sec. \ref{sec:4}.

\subsection{Radial equations for TT eigenmodes}

\par For the TT related radial equation, we set $\psi_{r}$ and $\psi_{t}$ in the same manner as in Eq.~(\ref{nttSVF}), and the angular component changes to
\begin{equation}
\begin{aligned}
\psi_{\theta_{i}}=\phi_{\theta}\otimes\bar{\psi}_{\theta_{i}}\; ,
\end{aligned}
\end{equation}
 where $\bar{\psi}_{\theta_{i}}$ is the TT mode eigenspinor-vector, with eigenvalue $\bar{\zeta}=j+(D-3)/2$, with $j=3/2,\,5/2,\,7/2,\,...$, as described in Ref.~\cite{ccch2016}. $\phi_{\theta}$ depends only on the radial coordinate $r$, and it behaves like a 2-spinor. We again use the Weyl gauge, where for the TT case this means $\phi_{t}=0$, and $\phi_{r}$ will automatically be zero to satisfy the equations of motion (as in our previous works, Refs.~\cite{ccch2016,ccch2018}). The only non-zero equation of motion is
\begin{equation}\label{TTEoM}
\begin{aligned}
\left(\frac{1}{r\sqrt{f}}i\sigma^{3}\partial_{t}+\frac{\sqrt{f}}{r}\sigma^{2}\partial_{r}+\frac{f'}{4r\sqrt{f}}\sigma^{2}+\frac{\sqrt{f}}{2r^{2}}(D-4)\sigma^{2}+\frac{i\bar{\zeta}}{r^{2}}\sigma^{1}+\frac{i\sqrt{\Lambda}}{r}\sqrt{\frac{D-2}{2(D-1)}} \right)\phi_{\theta}=0\; .
\end{aligned}
\end{equation}
In this case the function $\phi_{\theta}$ is already gauge invariant, and we set
\begin{equation}
\phi_{\theta}=f^{-\frac{1}{4}}r^{-\left(\frac{D-4}{2}\right)}\sigma^{2}e^{-i\omega t}\tilde{\phi}_{\theta} \; .
\end{equation}
We then obtain the equation
\begin{equation}\label{TTEoM}
\left(f\partial_{r}-\frac{\bar{\zeta}\sqrt{f}}{r}\sigma^{3}+\sqrt{\frac{f\Lambda \left(D-2\right)}{2\left(D-1\right)}}i\sigma^{2}\right)\tilde{\phi}_{\theta}=i\omega\sigma^{1}\tilde{\phi}_{\theta} \; .
\end{equation}
Following a similar method as in the non-TT case, we take
\begin{equation}\label{TTComp}
\tilde{\phi}_{\theta}=\left[\sin\left(\frac{\theta}{2}\right)\sigma^{3}+\cos\left(\frac{\theta}{2}\right)\sigma^{2}\right]
\begin{pmatrix}
\varphi_{1}\\
\varphi_{2}\\
\end{pmatrix} \; ,
\end{equation}
where $\varphi_{1,2}$ are functions of $r$ only and
\begin{equation}
\theta=\tan^{-1}\left(\frac{ir}{\bar{\zeta}}\sqrt{\frac{\Lambda\left(D-2\right)}{2\left(D-1\right)}}\right) \; .
\end{equation}
As a result, Eq.~(\ref{TTEoM}) can be rewritten into two coupled equations
\begin{equation}
\begin{aligned}
\left(\mathbb{F}\partial_{r}+\mathbb{W}\right)\varphi_{1} = -i\omega\varphi_{2}\; ,\\
\left(\mathbb{F}\partial_{r}-\mathbb{W}\right)\varphi_{2} = -i\omega\varphi_{1}\; ,
\end{aligned}
\end{equation}
where
\begin{eqnarray}\label{TTSP}
\mathbb{F}&=&f\left[1-\frac{if}{2\omega}\sqrt{\frac{\Lambda\left(D-2\right)}{2\left(D-1\right)}}\left(\frac{2\bar{\zeta}\left(D-1\right)}{2\bar{\zeta}^{2}\left(D-1\right)-r^{2} \Lambda\left(D-2\right)}\right)\right]^{-1}\; ,\nonumber \\
\mathbb{W}&=&\left(\frac{\bar{\zeta}^{2}f}{r^{2}}-\frac{f\Lambda\left(D-2\right)}{2\left(D-1\right)}\right)^{\frac{1}{2}} \left[1-\frac{if}{2\omega}\sqrt{\frac{\Lambda\left(D-2\right)}{2\left(D-1\right)}}\left(\frac{2\bar{\zeta}\left(D-1\right)}{2\bar{\zeta}^{2}\left(D-1\right)-r^{2} \Lambda\left(D-2\right)}\right)\right]^{-1}\; .\nonumber\\
\end{eqnarray}
Using the tortoise coordinate $\partial_{\hat{r}_{*}}=\mathbb{F}\partial_{r}$, we obtain the TT radial equation
\begin{equation}\label{TTRE}
\left(\partial^{2}_{\hat{r}_{*}}+\omega^{2}-\mathbb{V}_{eff}\right)\varphi_{1,2}=0\; ,
\end{equation}
where the effective potential is
\begin{equation}\label{TTEFV}
\mathbb{V}_{eff}=\mp\partial_{\hat{r}_{*}}\mathbb{W}+\mathbb{W}^{2}\; .
\end{equation}
We note that the TT related effective potential is basically constructed from the first terms of $\mathcal{B}$ and $\mathcal{D}$ in Eqs.~(\ref{NTTCOE}) of the non-TT related effective potential.
As another remark, we would like to point out that there are no TT eigenmodes on $S^{2}$, as the first one appears on $S^{3}$ \cite{ccch2016}. As such, we do not have the TT related radial equation in four dimensional cases, and Eq.~(\ref{TTRE}) can only be true for $D\geq5$.

%
%

\section{Properties of the effective potentials}\label{sec:4}

\par The effective potential is the most important physical quantity when working in black hole perturbation theory, as it determines the behaviors of different spin fields in a curved spacetime. We have shown in sec. \ref{sec:3} the non-TT related and TT related effective potentials of spin-3/2 fields in a general dimensional Schwarzschild (A)dS spacetime in Eqs.~(\ref{NTTEFV}) and (\ref{TTEFV}). Note that both of these potentials represent the behavior of spin-3/2 fields, and these potentials are analogous to both the Regge-Wheeler potential \cite{rg1957} and the Zerilli potential \cite{z1970}, which are related to the spin-2 fields in spherically symmetric black hole backgrounds. It is clear that the basic characteristics for the effective potentials in Eqs.~(\ref{NTTEFV}) and (\ref{TTEFV}) are that they both include an imaginary part and an $\omega$ dependence. This imaginary part and $\omega$ dependence are also present in the Teukolsky equations \cite{t1973}, where the background spacetime in that case is the Kerr black hole. As such, this is the first time when the background spacetime has been spherically symmetric and the effective potential contains both of these characteristics. Note that until this point the cosmological constant has been left general, that is, the effective potentials in Eqs.~(\ref{NTTEFV}) and (\ref{TTEFV}) represent both the asymptotically dS and AdS cases. We shall analyze our background further by chosing some specific parameters and studying how the effective potentials behave.

\subsection{Schwarzschild de Sitter cases}

\par In this case the cosmological constant $\Lambda\geq0$, and in general the imaginary part and the $\omega$ dependence remain for both the non-TT related and the TT related cases. In the earlier studies of the Teukolsky equation, Chandrasekhar and Detweiler \cite{cd1976,d1977} imposed a transformation on the radial equation to ensured it possessed a purely real and short range potential, later Sasaki and Nakamura extended this method for studying the radial equation with a source \cite{sn1982}. However, a direct application of the method in these references may not work for our effective potential, and some further analytic study is needed, which is beyond the scope of the current work.

\par
\begin{figure}
\begin{subfigure}{0.45\textwidth}
\includegraphics[width=\textwidth]{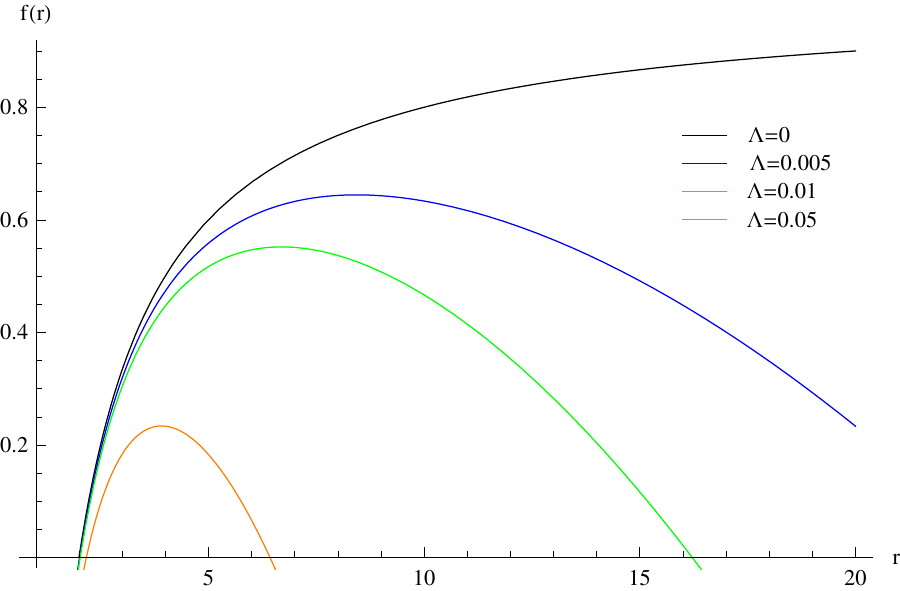}
\caption{The metric function $f(r)$.}\label{dSBHfr}
\end{subfigure}
\begin{subfigure}{0.45\textwidth}
\includegraphics[width=\textwidth]{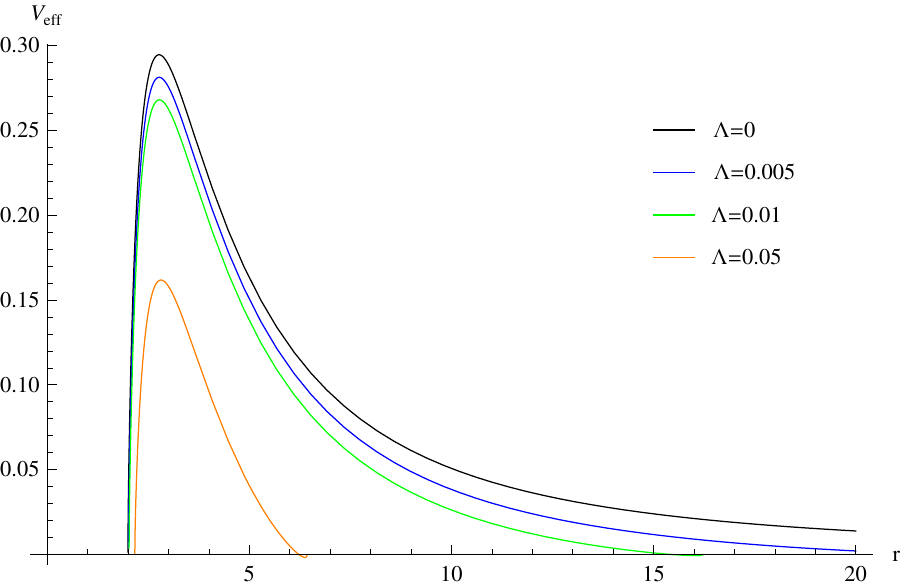}
\caption{The spin-3/2 non-TT effective potential in a de Sitter black hole spacetime.}\label{dSBHV}
\end{subfigure}
\caption{Plots of the metric function and non-TT effective potential for $D = 4$, $j = 5/2$ and $\Lambda = 0, 0.005, 0.01, 0.05$.}\label{dSBH}
\end{figure}

 Nevertheless, there is a special case for the non-TT related potential, when $D=4$. The fraction of $\mathcal{D}$ over $i\mathcal{B}$ in Eq.~(\ref{NTTSP}) becomes $r$ independent in this case, and as such the second term in the square parentheses in Eq.~(\ref{NTTSP}) vanishes. We may conclude that if $\mathcal{D}^{2}-\mathcal{B}^{2}\geq0$, the effective potential shall be purely real and without an $\omega$ dependence. The potential then behaves like a standard black hole perturbation theory case, which is barrier like and vanishes at the event horizon as well as the cosmic horizon. In Fig.~\ref{dSBH} we present the case of $M=1$ and $j=5/2$, where the corresponding changes in the metric function and the effective potential are shown when the cosmological constant $\Lambda$ is varied. Note that the $\Lambda=0$ case is equivalent to the Schwarzschild case studied previously \cite{ccch2016,ccchn2015}.

\subsection{Schwarzschild Anti de Sitter cases}

\par
\begin{figure}
\begin{subfigure}{0.45\textwidth}
\includegraphics[width=\textwidth]{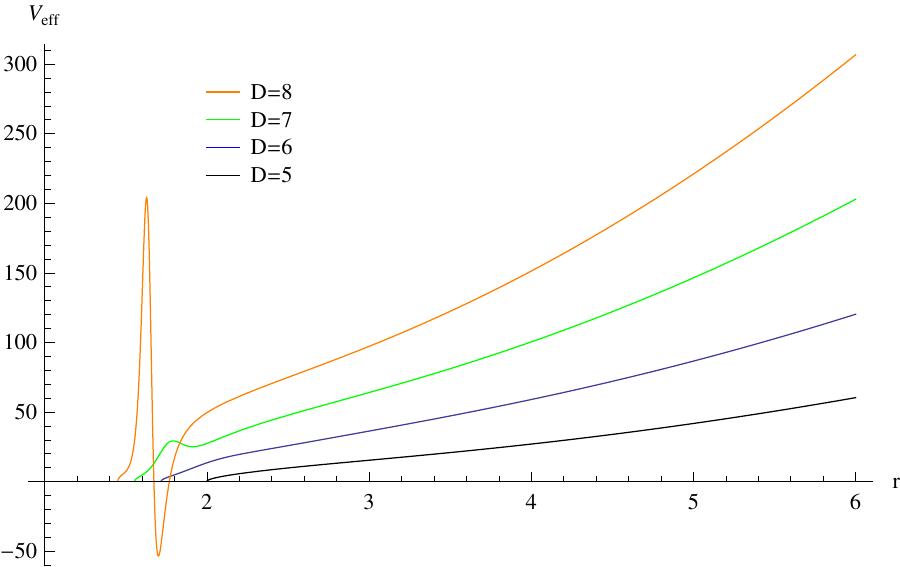}
\caption{For $M=1$.}\label{AdSVM1}
\end{subfigure}
\begin{subfigure}{0.45\textwidth}
\includegraphics[width=\textwidth]{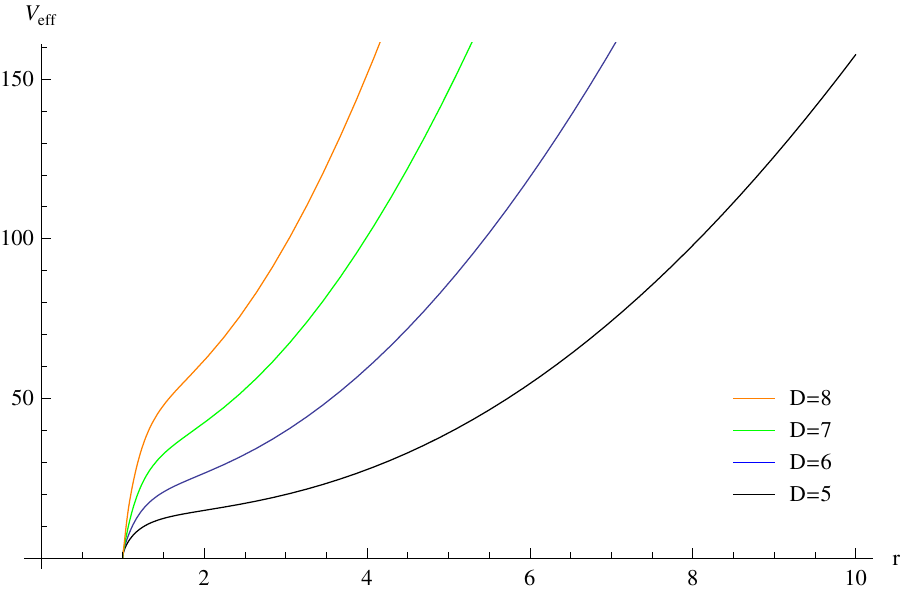}
\caption{For $M=10$.}\label{AdSVM10}
\end{subfigure}
\caption{The spin-3/2 non-TT effective potential in Anti de Sitter black hole spacetimes for $D=5,6,7,8$, $j=5/2$, $\Lambda = -\frac{(D-1)(D-2)}{2}$, and $\omega=10$.}\label{SAdSV}
\end{figure}

 For the effective potential in Eqs.~(\ref{NTTEFV}) and (\ref{TTEFV}) with cosmological constant $\Lambda\leq0$, we have found that the imaginary part automatically vanishes, leaving a purely real valued effective potential, where this can be easily deduced from the $\mathcal{D}$ function in Eq.~(\ref{NTTCOE}), the superpotential $W$ in Eq.~(\ref{NTTSP}) for the non-TT related cases, as well as from the superpotential $\mathbb{W}$ in Eq.~(\ref{TTSP}) for TT related cases. However, the $\omega$ dependence remains, and because of this dependence, we are not able to deduce the exact behavior of the effective potential until we solve the QNM frequency $\omega$. We can, nevertheless, still estimate the behavior of the potential, in particularly its asymptotic behavior, for some given value of $\omega$, as in Fig.~\ref{SAdSV}. Note that the parameter choice in these plots shall be $\Lambda = -(D-1)(D-2)/2$, which represents the ``AdS radius" as one unit, with $M$ set to 1 and then 10, and $\omega=10$, along with a variation of the dimension $D$ for the non-TT effective potential.

\par In the AdS spacetimes the standard expectation is that the potential function will tend to infinity as $r\rightarrow \infty$, and the fields will be reflected by this infinite potential. This is true for our cases when $D\geq5$, both for non-TT and TT modes. However, it is not true for the 4-dimensional case, which we shall analyze further in the next subsection. For the higher dimensional Schwarzschild Anti de Sitter cases, though, it is clearly the most suitable candidate for studying the QNMs, see sec.~\ref{sec:5}, as we can use a well known method in obtaining the QNM frequencies without too much modification \cite{hh2000}. 

\subsection{The 4-dimensional Schwarzschild Anti de Sitter case}

\par
\begin{figure}
\begin{subfigure}{0.45\textwidth}
\includegraphics[width=\textwidth]{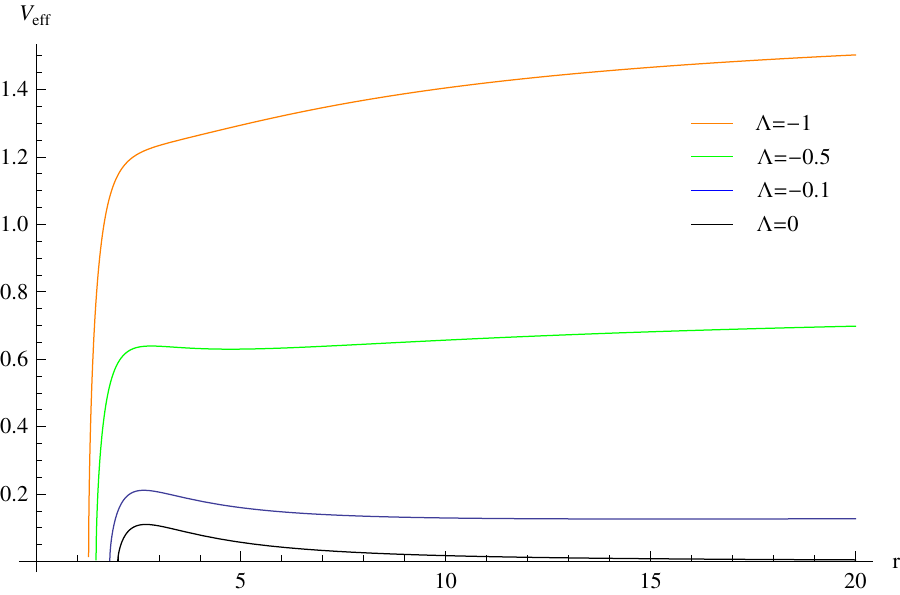}
\caption{For $j=3/2$ and $M=1$.}\label{AdSVM}
\end{subfigure}
\begin{subfigure}{0.45\textwidth}
\includegraphics[width=\textwidth]{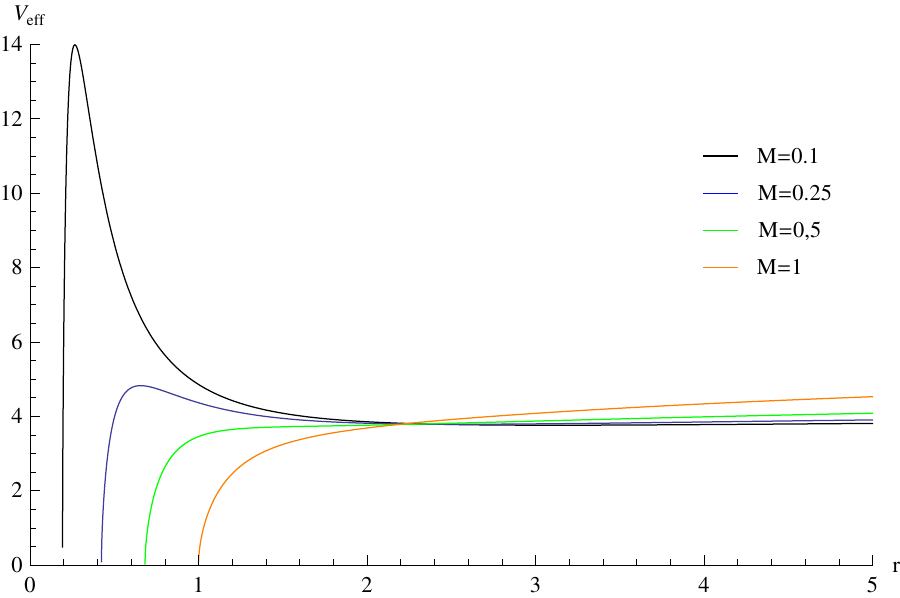}
\caption{For $j=3/2$ and $\Lambda=-3$ (taking the AdS radius as one).}\label{AdSVL}
\end{subfigure}
\caption{The spin-3/2 non-TT effective potential in the 4-dimensional Schwarzschild AdS spacetime.}\label{4dAdS}
\end{figure}

 In the 4-dimensional case the second term in the square parentheses of Eq.~(\ref{NTTSP}) vanishes (similarly in the 4-dimensional de Sitter case), and the effective potential becomes $\omega$ independent. Furthermore, we also have similar features to the higher dimensional Schwarzschild Anti de Sitter case, in that it becomes a purely real potential function. From these two points, the effective potential acquires the standard form of other previously studied fields in spherically symmetric black hole spacetimes (see, for example, Ref.~\cite{ccdn2007}). However, the usual infinite potential wall at infinity of the AdS black hole does not, in general, appear. Instead the potential tends to some finite non-zero value when $r\rightarrow \infty$. In Fig.~\ref{4dAdS} we show two sets of effective potentials with different parameter choices: In Fig.~\ref{AdSVM} we take the black hole mass as one, where the $\Lambda=0$ case represents the flat Schwarzschild case. In Fig.~\ref{AdSVL} we take the AdS radius as one and plot different values of the mass. Both of these plots present an evolution from a barrier like potential to a step like potential, and the potential reaches an asymptotic finite non-zero value that becomes larger for larger values of $|\Lambda|$ or $M$. From these two plots, we can also see that for small AdS black holes ($\Lambda$ small or $M$ small), the potential possesses a barrier shape near the event horizon, while for large black holes, this barrier disappears and the potential becomes step like.

%
%

\section{Quasi-normal modes for asymptotically AdS cases}\label{sec:5}

\par From our discussions in sec. \ref{sec:4} on the effective potentials, we shall use the asymptotically AdS cases as an example for studying the QNMs using the Horowitz-Hubeny method \cite{hh2000}. The Horowitz-Hubeny method works by taking a Taylor series expansion of the potential function around the event horizon, and then imposing a purely ingoing boundary condition near the event horizon, and Dirichlet boundary conditions on the field solutions at spatial infinity. However, the Taylor series expansion for our potential will include non-integer orders. As such, modifications are necessary for this method to be applied in the current situation.

\subsection{The modified Horowitz-Hubeny method}

\par To investigate the entire region $r_{+}<r<\infty$, where $r_{+}$ is the event horizon, we map the space to a finite region by changing coordinates to $x=1/r$. We begin by setting the purely ingoing boundary condition $\phi_{2} \sim \Xi(r)\,e^{-i\omega r_{*}} $, and the radial equation, Eq.~(\ref{NTTRD}),  becomes
\begin{equation}
\left[\mathcal{F}\frac{d^{2}}{dr^{2}} +\left(\mathcal{F}'-2i\omega \right)\frac{d}{dr} - \left(\frac{V_{eff}}{\mathcal{F}} \right) \right]\Xi(r)=0 \; .
\end{equation}
By redefining $\Lambda = - (D-1)(D-2)/2$, the coefficient of the $r^{2}$ term in $f(r)$ becomes $-1$, which is equivalent to saying that we are setting the AdS radius to one. Using $r=1/x$, and 
$$M=\frac{1}{2}\left[\frac{1+x_{+}^{2}}{x_{+}^{(D-1)}} \right] \; ,$$
where $x_{+}=1/r_{+}$ and $f(r_{+})=0$, the radial equation becomes
\begin{equation}\label{EQ:HHPot}
\left[s(x)\frac{d^{2}}{dx^{2}} + \frac{t(x)}{(x-x_{+})}\frac{d}{dx} - \frac{u(x)}{(x-x_{+})^{2}} \right]\Xi(x)=0 \; ,
\end{equation}
with
\begin{equation}\label{EQ:TermsInEquations}
\begin{aligned}
s(x)&=\frac{\mathcal{F}\left(x\right)}{x-x_{+}}x^{4}\; ,\\
t(x) &= 2x^{3}\mathcal{F}\left(x\right)+x^{4}\left(\partial_{x}\mathcal{F}\left(x\right)\right) +2x^{2}i\omega\; ,\\
u(x) &= (x-x_{+})\left(\frac{V_{eff}}{\mathcal{F}\left(x\right)} \right)\; .
\end{aligned}
\end{equation}
In order for $\Xi(x)$ to satisfy the in-going boundary condition at the horizon, we must write it as
\begin{equation}\label{EQ:SeriesExpPhi}
\Xi(x)=\sum\limits_{k=0}^{n}a_{\frac{k}{2}}\left(x-x_{+}\right)^{\frac{k}{2}}\; .
\end{equation}
Note that half-integral powers $k/2$ are necessary, as we require non-integer terms in the Taylor expansion of $u(x)$, where $n/2$ is the order of the expansion, and $n$ shall be an even integer for ease of calculation. Plugging Eq.~(\ref{EQ:SeriesExpPhi}) into Eq.~(\ref{EQ:HHPot}), and expanding Eq.~(\ref{EQ:TermsInEquations}) when $x$ is close to $x_{+}$, we can then compare $(x-x_{+})$ order by order to determine the value of $a_{\frac{k}{2}}$. That is,
\begin{eqnarray}\label{coe}
a_{\frac{k}{2}}&=&-\frac{1}{P_{\frac{k}{2}}}\sum\limits_{q=0}^{k-1}\left[\frac{q}{2}\left(\frac{q}{2}-1\right)s_{\frac{k-q}{2}}+\frac{q}{2}t_{\frac{k-q}{2}}-u_{\frac{k-q}{2}}\right]a_{\frac{q}{2}} \; ,\nonumber\\
P_{\frac{k}{2}}&=&\frac{k}{2}\left(\frac{k}{2}-1\right)s_{0}+\frac{k}{2}t_{0} \; ,
\end{eqnarray}
where $k=0, 1, 2,...n$, $s_{\frac{k}{2}}$, $t_{\frac{k}{2}}$, and $u_{\frac{k}{2}}$ represent the corresponding coefficients of the expansion polynomials.

\par We next consider the Dirichlet boundary condition, which means that the wave function must vanish when $x\rightarrow 0$, therefore we have
\begin{equation}\label{Dirich}
\Phi=\sum\limits_{k=0}^{n}a_{\frac{k}{2}}\left(-x_{+}\right)^{\frac{k}{2}}=0 \; .
\end{equation}
Putting all the expressions in Eq.~(\ref{coe}) into Eq.~(\ref{Dirich}), allows us to solve the QNM frequencies with respect to the $\frac{n}{2}$-th order expansion. As a remark, note that the notation shown in this subsection corresponds to the non-TT related cases only. However, for the TT related cases, just a replacement of $\mathcal{F}$ and $V_{eff}$ by $\mathbb{F}$ and $\mathbb{V}_{eff}$ is needed.

\subsection{Quasi-normal modes and some related difficulties}

\par In the numerical calculation of the QNM frequencies in the currently considered cases, there were some intrinsic difficulties, primarily related to computing power. For the scalar perturbation case of Horowitz and Hubeny \cite{hh2000}, their method gave stable results to three significant figures with an expansion to around a hundred orders. These results were reproducible by us after several hours of computing time for one mode. Note that a large number of studies for bosonic perturbations \cite{kono2002,cl2001,ckl2003,wlm2004,whs2015} and Dirac perturbations \cite{gj2005,whj2017} using this method have been performed, and we suppose that these calculations were also working to a similar order of expansion. However, due to the complexity of our spin-3/2 effective potential, an expansion to a hundred orders in our current case would be much more difficult to achieve.

\par It is well known that bosonic perturbations in all types of spherically symmetric spacetimes are determined by the Regge-Wheeler equation \cite{rg1957} and the Zerilli equation \cite{z1970}, which are not greatly different with regards to performing numerical calculations. For the other case of Dirac perturbations \cite{ccdn2007}, the super potential related to the effective potential shall be $\lambda\sqrt{f}/r$, which is just the leading term of our coefficient $\mathcal{B}$ in Eq.~(\ref{NTTCOE}), this is also true for our TT related cases. Together with our earlier discussions, as related to the $\omega$ dependence and the non-integer order required when doing the Taylor expansions, we are faced with a far more complicated and nested algorithm that would take a lot of computing time to converge to a reliable QNM frequency.

\par As such, we present the QNM frequencies from this Horowitz-Hubeny method with expansions up to the $14^{th}$ order. To achieve results to this order already required more than 10 hours per mode, and to go to the next order would cost approximately three times more computing time. In the following tables, Tabs. \ref{Tab1}, \ref{Tab2}, and \ref{Tab3}, we present the results for the QNMs of the non-TT related potential for the cases of $D=5, 6, 7 ,8$, $x_{+}=0.001, 0.01, 0.025, 0.1, 1, 1.2$, and $\ell=0, 1, 2$, where $\ell=j-3/2$ is related to the spinor-vector eigenmodes on $S^{N}$. 

\par
\begin{figure}
\begin{subfigure}{0.45\textwidth}
\includegraphics[width=\textwidth]{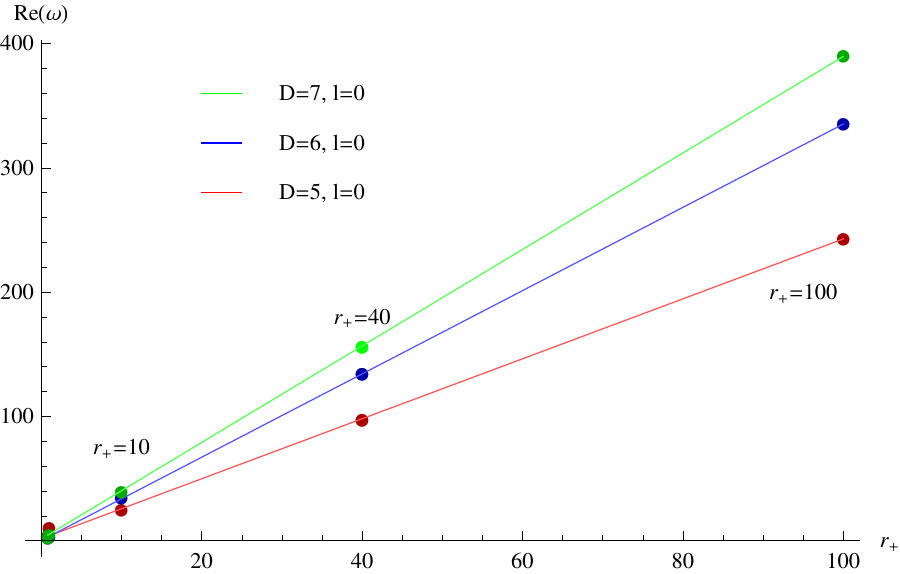}
\caption{Linear relation for the real part up to the $14^{th}$ iteration.}
\end{subfigure}
\begin{subfigure}{0.45\textwidth}
\includegraphics[width=\textwidth]{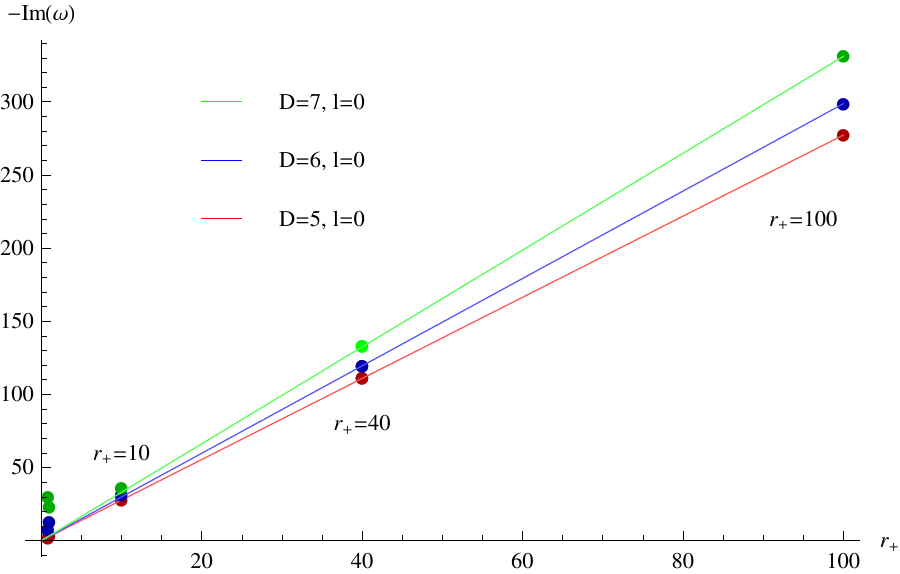}
\caption{Linear relation for the imaginary part up to the $14^{th}$ iteration.}
\end{subfigure}
\caption{QNMs, non-TT related case, with $\ell=0$, $D=5,6,7$, and $x_{+}=0.01, 0.025, 0.1, 1, 1.2$.}\label{QNMsplot1}
\end{figure}

 In Fig.~\ref{QNMsplot1}, we have plotted the real and the imaginary parts of the QNM frequencies separately, for the non-TT case with $\ell=0$, versus the position of the event horizon $r_{+}$. Both the real parts and the magnitudes of the imaginary parts of the frequencies increase with $r_{+}$, as well as with the spacetime dimension $D$. Moreover, for large black holes ($r_{+}=100, 40, 10$ or $x_{+}=0.01, 0.025, 0.1$), a linear relation between the QNM frequencies (${\rm Re}(\omega)$ or $|{\rm Im}(\omega)|$) and $r_{+}$ is apparent. This is consistent with the study of Horowitz and Hubeny \cite{hh2000} in which they found these linear relations between QNM frequencies and Hawking temperatures of large AdS black holes, as the temperature is proportional to $r_{+}$ for large AdS black hole cases. Note that for small black holes ($r_{+}=1, 0.83$ or $x_{+}=1, 1.2$), as indicated in Fig.~4, the QNM frequencies do not conform to this relation. Lastly, from Tabs. I, II, and III, we can see that the values of the frequencies do not vary with the angular momentum parameter $\ell$. Hence, we expect the linear relation to hold also for large black holes in the $\ell=1, 2$ cases.

\begin{table}
\centering
\caption{Spin-3/2 non-TT related QNMs for $\ell=0$ in $D=5,6,7,8$-dimensional Schwarzschild AdS black hole spacetimes.}
\begin{tabular}{| c | c | c | c | c | }
\hline
 & \multicolumn{4}{|c|}{QNMs} \\
\hline
$x_{+}$ & $D=5$ & $D=6$ & $D=7$ & $D=8$ \\
\hline
\hline
0.001 & 2421.34 -2768.09 i &  3345.87 -2982.35 i &  3892.93 -3306.22 i & 4260.42 -3581.29 i \\
0.01  & 242.6 -277.064 i   &  335.012 -298.455 i &  389.523 -330.907 i & 425.945 -358.546 i \\
0.025 & 97.3393 -110.99 i  &  134.267 -119.606 i &  155.722 -132.922 i & 169.601 -144.366 i \\
0.1   & 24.7773 -28.0351 i &  34.051 -30.9378 i  &  38.8121 -36.1762 i & 50.6502 -10.5606 i \\
1     & 3.97854 -2.91456 i &  2.92626 -13.028 i  &  4.6083 -23.1799 i  & 8.95824 -13.0857 i \\
1.2   & 1.69155 -0.02631 i &  2.41208 -6.97679 i &  1.95646 -29.6368 i & 8.22353 -25.1656 i \\
\hline
\end{tabular}
\label{Tab1}
\end{table}

\begin{table}
\centering
\caption{Spin-3/2 non-TT related QNMs for $\ell=1$ in $D=5,6,7,8$-dimensional Schwarzschild AdS black hole spacetimes.}
\begin{tabular}{| c | c | c | c | c | }
\hline
 & \multicolumn{4}{|c|}{QNMs, $\ell=1$} \\
\hline
$x_{+}$ & $D=5$ & $D=6$ & $D=7$ & $D=8$ \\
\hline
0.001 & 2421.55 -2768.2 i  & 3346.03 -2982.42 i &  3893.03 -3306.27 i &  4260.47 -3581.34 i  \\
0.01  & 242.779 -277.174 i & 335.142 -298.591 i &  389.416 -331.262 i &  425.4 -359.119 i  \\
0.1   & 25.0869 -28.3967 i & 34.6927 -32.1749 i &  57.1463 -13.584 i  &  59.2126 -9.43011 i  \\
1     & 4.02362 -1.92365 i & 2.43502 -6.05751 i &  1.74403 -15.59 i   &  2.01607 -28.2032 i  \\
1.2   & 3.91933 -0.81368 i & 0.81334 -3.40948 i &  1.47253 -13.8613 i &  1.70049 -25.9925 i  \\
\hline
\end{tabular}
\label{Tab2}
\end{table}

\begin{table}
\centering
\caption{Spin-3/2 non-TT related QNMs for $\ell=2$ in $D=5,6,7,8$-dimensional Schwarzschild AdS black hole spacetimes.}
\begin{tabular}{| c | c | c | c | c | }
\hline
 & \multicolumn{4}{|c|}{QNMs, $\ell=2$} \\
\hline
$x_{+}$ & $D=5$ & $D=6$ & $D=7$ & $D=8$ \\
\hline
\hline
0.001 & 2421.76 -2768.32 i  & 3346.19 -2982.49 i  &  3893.13 -3306.32 i &  4260.5 -3581.41 i \\
0.01  & 242.947 -277.283 i  & 335.263 -298.753 i  &  389.253 -331.712 i &  424.717 -359.844 i  \\
0.1   & 25.224 -28.9174 i   & 34.3238 -32.525 i   &  40.9097 -38.3034 i &  46.9668 -44.8927 i  \\
1     & 1.51351 -0.0204 i   & 4.69729 -0.676194 i &  5.985 -10.5436 i   &  1.04004 -17.6066 i  \\
1.2   & 0.989231 -0.0587 i  & 1.25176 -0.20764 i  &  4.43287 -6.24447 i &  0.787022 -9.37979 i  \\
\hline
\end{tabular}
\label{Tab3}
\end{table}

%
%

\section{Conclusion}\label{sec:6}

\par The main result of this paper is the obtaining of the master radial equations of spin-3/2 fields in the Schwarzschild (A)dS black hole spacetimes. This has been done for the non-TT related case in Eq.~(\ref{NTTEFV}), and the TT related case in Eq.~(\ref{TTEFV}). A systematic study of the effective potentials has also been performed. However, due to the varied complexities of these potentials, the issues around calculating the QNMs for asymptotic dS spacetimes, and the study of a special case for the 4-dimensional Schwarzschild AdS black hole, must be left for future works as greatly different techniques for their calculation need to be explored and developed. This being something beyond the focus of the current investigation of obtaining the master radial equations for spin-3/2 fields in Schwarzschild (A)dS spacetimes. Nevertheless, we have presented preliminary results for the QNMs in higher dimensional asymptotic AdS black hole cases, where it was still possible to observe a linear relationship between QNM frequencies and the corresponding Hawking temperatures of the large AdS black holes, as previously noted in scalar perturbation studies \cite{hh2000}. 

\begin{figure}
\begin{subfigure}{0.45\textwidth}
\includegraphics[width=\textwidth]{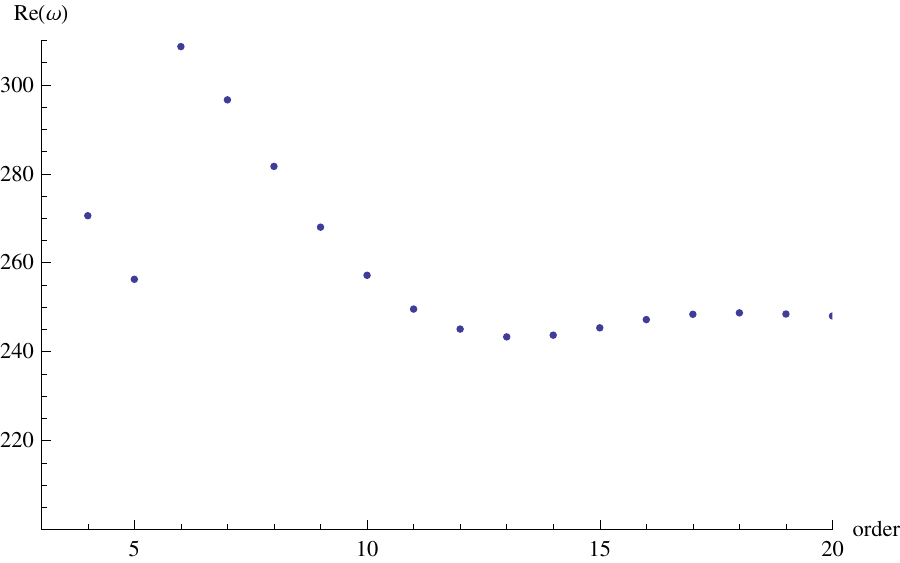}
\caption{Real part.}
\end{subfigure}
\begin{subfigure}{0.45\textwidth}
\includegraphics[width=\textwidth]{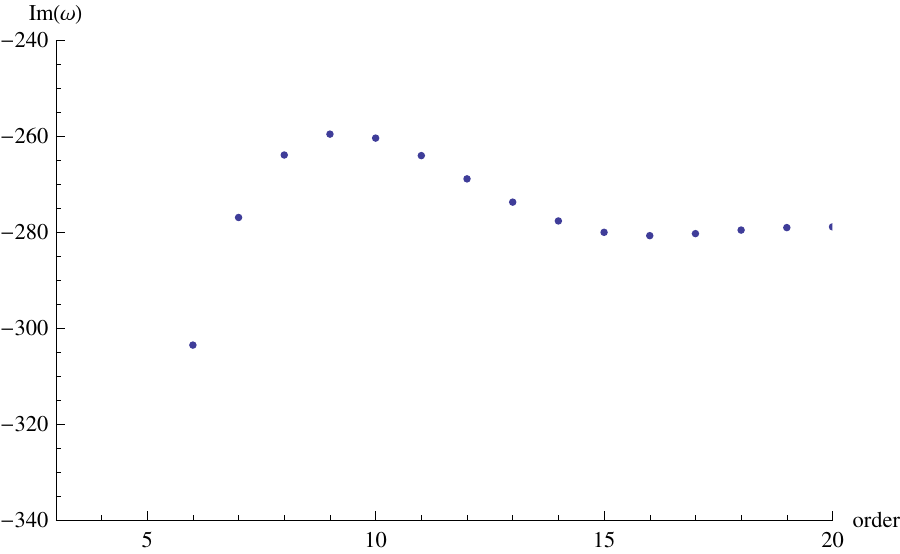}
\caption{Imaginary part.}
\end{subfigure}
\caption{QNMs, TT related case with $20^{th}$ order expansion, $D=5$, $r_{+}=100$, $\ell=0$.}\label{QNMSplot2}
\end{figure}

To have an idea on the accuracy of the values of the QNM frequencies in our tables, we present in Fig.~\ref{QNMSplot2} a plot for the convergence of the QNMs up to $20^{th}$ order in the expansion of a mode of the TT related potential. As we mentioned earlier, the TT related effective potential is just the leading term of the non-TT related effective potential, and as such has allowed us to extend these numerics to a few higher orders. The values in Fig.~\ref{QNMSplot2} start to stabilize after 15 or 16 orders of expansion. Hence, this indicates that the values of the QNM frequencies in Tabs. I, II and III can  be taken as, at best, approximate. In order to obtain more accurate values of the QNM frequencies, as well as those of higher overtone QNMs, the expansion in the Horowitz-Hubeny method would need to be taken to much higher orders. As noted earlier, the computer power and time required to do this is not feasible, and as such a new numerical method, for example the ``asymptotic iteration method'' \cite{ccdhn2012}, would be required to evaluate the QNM frequencies in these cases to a more desirable accuracy. However, as noted for the dS and 4-dimensional cases, such numerical methods require a more dedicated and detailed investigation beyond the scope of this current study.


%
%

\section*{Acknowledgements}

ASC and GEH are supported in part by the National Research Foundation of South Africa. HTC is supported in part by the Ministry of Science and Technology, Taiwan, ROC under the Grant Nos.~MOST106-2112-M-032-005 and MOST107-2112-M-032-011. HTC is also
supported in part by the National Center for Theoretical Sciences (NCTS).

%
%

\end{document}